\documentclass[journal]{IEEEtran}
\usepackage{amsmath,amssymb,amsfonts,amsthm}
\usepackage{algorithm,algorithmic}
\usepackage{cite}
\usepackage{array}
\usepackage[caption=false,font=normalsize,labelfont=sf,textfont=sf]{subfig}
\usepackage{textcomp}
\usepackage{stfloats}
\usepackage{url}
\usepackage{verbatim}
\usepackage{graphicx}
\usepackage[dvipsnames, svgnames, x11names]{xcolor}

\newtheorem{Lemma}{Lemma}

\def\BibTeX{{\rm B\kern-.05em{\sc i\kern-.025em b}\kern-.08em
    T\kern-.1667em\lower.7ex\hbox{E}\kern-.125emX}}
\usepackage{balance}
\usepackage[colorlinks=true,linkcolor=red,citecolor=green]{hyperref}
\title{Weighted Sum-Rate Maximization for Movable Antenna-Enhanced Wireless Networks}
\begin{document}

\author{
\IEEEauthorblockN{Biqian Feng, Yongpeng Wu, \emph{Senior Member, IEEE}, Xiang-Gen Xia, \emph{Fellow, IEEE}, \\ and Chengshan Xiao, \emph{Fellow, IEEE}}
\thanks{B. Feng and Y. Wu are with the Department of Electronic Engineering, Shanghai Jiao Tong University, Minhang 200240, China (e-mail: fengbiqian@sjtu.edu.cn; yongpeng.wu@sjtu.edu.cn). }
\thanks{X.-G. Xia is with the Department of Electrical and Computer Engineering, University of Delaware, Newark, DE 19716, USA. (e-mail: xxia@ee.udel.edu).}
\thanks{C. Xiao is with the Department of Electrical, and Computer Engineering, Lehigh University, Bethlehem, PA 18015 USA (e-mail: xiaoc@lehigh.edu).}
\thanks{Corresponding author: Yongpeng Wu.}
}

\maketitle
\begin{abstract}
This letter investigates the weighted sum rate maximization problem in movable antenna (MA)-enhanced systems. To reduce the computational complexity, we transform it into a more tractable weighted minimum mean square error (WMMSE) problem well-suited for MA. We then adopt the WMMSE algorithm and majorization-minimization algorithm to optimize the beamforming and antenna positions, respectively. Moreover, we propose a planar movement mode, which constrains each MA to a specified area, we obtain a low-complexity closed-form solution. Numerical results demonstrate that the MA-enhanced system outperforms the conventional system. Besides, the computation time for the planar movement mode is reduced by approximately 30\% at a little performance expense.
\end{abstract}

\section{Introduction}
With the advancement of multiple-input multiple-output (MIMO) technologies, wireless communication systems have witnessed a substantial increase in capacity due to the utilization of additional spatial degrees of freedom (DoFs). However, the conventional MIMO systems equipped with fixed-position antennas (FPAs) face challenges in further enhancing performance. Since the last century, the concept of antenna arrays with arbitrarily distributed elements \cite{NUAA} has gradually surfaced. This concept suggests that fewer, non-uniformly deployed elements can match the performance as of equally spaced arrays. Recently, with the gradual maturation of micro-electromechanical systems (MEMS), and liquid metal technologies, movable antenna (MA) \cite{Wenyan_MIMO}, also known as fluid antenna (FA) \cite{KKWong_FA}, flexible antenna \cite{Zijian_FA}, and flexible-position MIMO \cite{Jiakang_FP}, has been proposed to optimize physical antenna placement adaptively according to the time-varying dynamic of wireless channels to enhance MIMO system performance.

Several studies have emerged to analyze and optimize the MA-enhanced MIMO systems. For instance, in \cite{Hao_FAS_Channel_Estimation, Zhenyu_Channel_Estimation}, the authors focused on the channel estimation problem and utilized compressed sensing-based methods to efficiently estimate the angles of departure, angles of arrival, and complex coefficients of the multi-path components, which ensures the MA systems to reap performance gains. Also, based on the estimated channel, \cite{Lipeng_Analysis} analyzed the maximum channel gain achieved by a single receive MA as compared to its FPA counterpart, and \cite{Guojie_CoMP} leveraged the advantages of MA to improve signal-to-noise ratio (SNR) of multiple destinations for jointly decoding a common message. Currently, most research focused on rate maximization \cite{Xintai,Yuqi_FAS_Statistical_CSI,Zhenqiao_Sum_Rate,Zhenyu_Multiuser,Hao_FAS} and power control \cite{Yifei_Discrete,Haoran_Downlink}. In \cite{Xintai,Yuqi_FAS_Statistical_CSI}, the transceiver equipped with MAs was considered to enhance the performance of point-to-point MIMO systems. Furthermore, in \cite{Yifei_Discrete,Zhenqiao_Sum_Rate,Zhenyu_Multiuser,Hao_FAS}, the authors considered a multiuser transmission scenario where only the BS is equipped with MAs. Specifically, limited by the realistic electromechanical devices, the motions of the MA elements were modeled as discrete movements in \cite{Yifei_Discrete}. In addition, in \cite{Haoran_Downlink}, the authors considered a multiuser transmission scenario where only users are equipped with MAs. However, most works employ the gradient descent algorithm and continuously try to obtain a suitable step size satisfying the constraints, so the computational complexity is relatively high.

In this letter, we investigate a popular weighted sum-rate (WSR) maximization problem in an MA-enhanced multiuser MIMO system, where both BS and users are equipped with MAs. The main contributions are summarized as follows:
\begin{itemize}
\item We formulate a WSR maximization problem and transform it into a more tractable weighted sum mean-square error minimization problem that is compatible with MAs.

\item We employ the block coordinate descent (BCD) method to optimize all variables alternately and adopt a planar movement mode to achieve a closed-form solution for optimizing antenna positions, thereby efficiently reducing the computational complexity.

\item Numerical results demonstrate that the MA-enhanced systems ensure superior performance over the conventional systems. Besides, the planar movement mode significantly reduces the computational complexity at a little performance expense.
\end{itemize}

\textit{Notations:} We adopt $x$, $\mathbf x$, and $\mathbf X$ to denote a scalar, vector, and matrix, respectively. 
Superscripts $*, T$, and $H$ stand for the conjugate, transpose, and conjugate transpose, respectively. $\Pi_{\mathcal C}(\mathbf x)$ denotes the projection of $\mathbf x$ onto the set $\mathcal C$. $\odot$ denotes the Hadamard product.

\section{System Model}
\begin{figure}[t]
\centering
\includegraphics[width=1\linewidth]{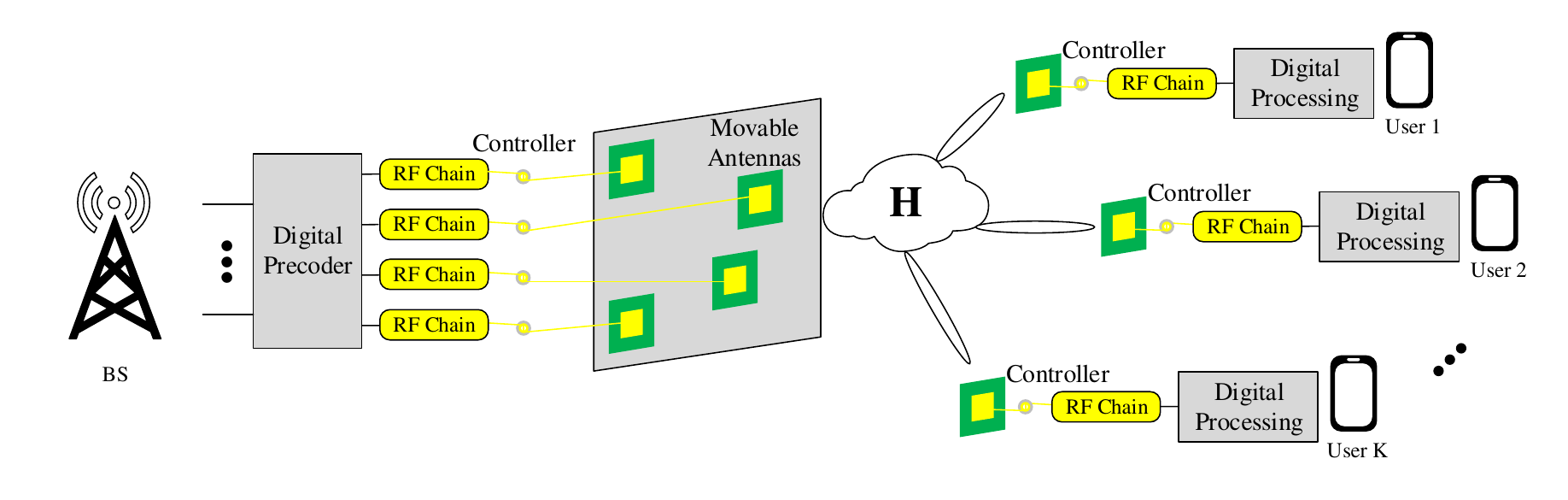}
\caption{System model: an MA-enhanced multiuser MIMO system.}
\label{syestem_model}
\end{figure}
As shown in Fig.~\ref{syestem_model}, we consider a multiuser downlink wireless communication system, where the BS equipped with $M$ MAs serves $K$ single-MA users. Compared with the conventional communication system, the position of each MA in the transmit region $\mathcal C^t$ and the receive region $\mathcal C_k^r, k=1,\cdots,K$, can be adjusted to exploit the spatial diversity gain via a controller, e.g., stepper motors \cite{Wenyan_MIMO} or liquid metal \cite{KKWong_FA}. 

Let $\mathbf t_m = (x_m, y_m)^T$ and $\mathbf r_k = (x_k, y_k)^T$ denote the position of MA $m$ at the BS and the position of the MA at user $k$, respectively, which are established by the classical Cartesian coordinate system. By stacking the positions of all antennas together, we denote the position vectors of the BS and users as $\mathbf t=(\mathbf t_1^T, \cdots, \mathbf t_M^T)^T$ and $\mathbf r=(\mathbf r_1^T, \cdots, \mathbf r_K^T)^T$, respectively. To avoid potential electrical coupling between adjacent MAs and upgrade the antenna efficiency in practical implementation, a minimum inter-MA distance $D$ is required between each pair of MAs at BS \cite{Wenyan_MIMO}, i.e., 
\begin{equation}
\label{minimum inter-MA distance}
\|\mathbf t_i - \mathbf t_j\|\geq D, \forall i \neq j.
\end{equation}
For the MA-enhanced system, the channel response depends on the position vector and the channel vector between the BS and user $k$ can be represented by a function of the position vector, i.e., $\mathbf h_k(\mathbf t, \mathbf r_k)\in\mathbb C^{M\times 1}$. Hence, the discrete-time signal received at user $k$ is written as
\begin{equation}
\label{received signal}
y_k = \mathbf h_k^H(\mathbf t,\mathbf r_k) \sum_{k=1}^K\mathbf w_k  s_k + n_k,
\end{equation}
where $\mathbf w_k\in\mathbb C^{M\times 1}$ and $s_k\in\mathbb C$ represent the beamformer and the unit-power complex valued information symbols intended for user $k$, respectively; $n_k\sim\mathcal{CN}(0,\sigma^2)$ models the additive noise with zero mean and variance $\sigma^2$. Based on \eqref{received signal}, the signal-to-interference-plus-noise ratio (SINR) experienced at user $k$ is obtained as
\begin{equation}
	\gamma_k\triangleq\frac{|\mathbf h_k^H(\mathbf t,\mathbf r_k) \mathbf w_k|^2}{\sum\limits_{i\neq k} |\mathbf h_k^H(\mathbf t,\mathbf r_k) \mathbf w_i|^2 + \sigma^2}.
\end{equation}

\subsection{Field-Response-Based Channel Model}
Let $L_k^t$ and $L_k^r$ denote the total numbers of transmit and receive channel paths from the BS to user $k$. Then, the signal propagation phase difference of the $l$-th transmit path for user $k$ between the $m$-th MA and the reference point $\mathbf t_0=[0,0]^T$ at the BS and the $l$-th receive path between the MA and reference point $\mathbf r_0=[0,0]^T$ at user $k$ are, respectively, given by
\begin{equation}
\begin{aligned}
\rho_{l,k}^t(\mathbf t_m) = \mathbf t_m^T\mathbf n_{l,k}^t  &= x_m\sin\theta_{l,k}^t\cos\phi_{l,k}^t + y_m\cos\theta_{l,k}^t,\\
\rho_{l,k}^r(\mathbf r_k) = \mathbf t_m^T\mathbf n_{l,k}^r &= x_k\sin\theta_{l,k}^r\cos\phi_{l,k}^r + y_k\cos\theta_{l,k}^r,
\end{aligned}
\end{equation}
where $\mathbf n_{l,k}^t\triangleq (\sin\theta_{l,k}^t\cos\phi_{l,k}^t, \cos\theta_{l,k}^t)^T$ and $\mathbf n_{l,k}^r\triangleq (\sin\theta_{l,k}^r\cos\phi_{l,k}^r, \cos\theta_{l,k}^r)^T$; $(\theta_{l,k}^t, \phi_{l,k}^t)$ and $(\theta_{l,k}^r, \phi_{l,k}^r)$ are the elevation and azimuth AoAs for the $l$-th transmit and receive path between user $k$ and the BS, respectively. Accordingly, the field-response vector of the receive channel paths between user $k$ and the $m$-th MA at the BS is given by \cite{Lipeng_Analysis}
\begin{equation}
\begin{aligned}
\label{eq: field-response vector}
\mathbf f_k(\mathbf t_m)&=\left[e^{j\frac{2\pi}{\lambda}\rho_{1,k}^t(\mathbf t_m)},\cdots, e^{j\frac{2\pi}{\lambda}\rho_{L_k^t,k}^t(\mathbf t_m)}\right]^T,\\
\mathbf g_k(\mathbf r_k)&=\left[e^{j\frac{2\pi}{\lambda}\rho_{1,k}^r(\mathbf r_k)},\cdots, e^{j\frac{2\pi}{\lambda}\rho_{L_k^r,k}^r(\mathbf r_k)}\right]^T.
\end{aligned}
\end{equation}
As a result, the channel vector between user $k$ and the BS is given by
\begin{equation}
\mathbf h_k(\mathbf t,\mathbf r_k) = \mathbf F_k^H(\mathbf t)\mathbf\Sigma_k\mathbf g_k(\mathbf r_k),
\end{equation}
where $\mathbf F_k(\mathbf t)\triangleq(\mathbf f_k(\mathbf t_1), \cdots, \mathbf f_k(\mathbf t_M))\in\mathbb C^{L_k\times M}$ denotes the field-response matrix at the BS, and $\mathbf \Sigma_k\in\mathbb C^{L_k^t \times L_k^r}$ is the path-response matrix, which represents the multi-path response between all the transmit and receive channel paths. As can be observed, the movement introduced by MA causes more significant variations of phases than the amplitudes \cite{Zhenyu_Multiuser}. 

\subsection{Problem Formulation}
In this letter, we aim to maximize the weighted sum rate of all the users by jointly designing the transmit beamformer and the antenna position vectors at both the BS and the $K$ users, subject to the minimum inter-MA distance \eqref{minimum inter-MA distance} at the BS and the transmit power constraints. Let $\mathbf W\triangleq(\mathbf w_1, \mathbf w_2, \cdots, \mathbf w_K) \in\mathbb C^{M\times K}$. The WSR maximization problem is formulated as
\begin{subequations}
\label{p: original problem}
\begin{align}
\underset{\mathbf W, \mathbf t,\mathbf r}{\text{maximize}} \quad& \sum\limits_{k=1}^K \alpha_k\log(1+\gamma_k) \label{p: original problem-a}\\
\text { subject to } &\sum\limits_{k=1}^K\|\mathbf w_k\|^2\leq P_\mathrm{max},\label{p: original problem-b}\\
&\|\mathbf t_i - \mathbf t_j\|_2\geq D, \forall i\neq j,\label{p: original problem-c}\\
&\mathbf t_i\in\mathcal C^t,\quad\mathbf r_k \in\mathcal C_k^r,\label{p: original problem-d}
\end{align}
\end{subequations}
where the weight $\alpha_k\geq 0$ represents the priority of user $k$. $P_\mathrm{max}$ denotes the maximum transmit power. The problem is challenging mainly due to the nonconvexity of the objective function \eqref{p: original problem-a} and the constraints \eqref{p: original problem-c}, which is an obstacle for the development of an optimal solution. To reduce the complexity and transform it into a more tractable problem, we consider the linear receive beamforming strategy so that the estimated signal at user $k$ is given by
\begin{equation}
\hat s_k = u_k^* y_k,
\end{equation}
where $\mathbf u\triangleq(u_1,u_2,\cdots,u_K)^T\in\mathbb C^{K\times 1}$ represents the receive beamformer. We assume the independence of $s_k$ and $n_k$. Then, the expected mean-square error (MSE) can be written as
\begin{equation}
\begin{aligned}
&e_k\triangleq\mathbb E\left[|\hat s_k - s_k|^2\right]\\
&=1 + |u_k|^2(\sigma^2 + \sum_{j=1}^K |\mathbf h_k^H(\mathbf t,\mathbf r_k)\mathbf w_j|^2) \\
&\quad\quad- 2\mathrm{Re}(u_k^* \mathbf h_k^H(\mathbf t,\mathbf r_k)\mathbf w_k).
\end{aligned}
\end{equation}
where $\mathrm{Re}(\cdot)$ denotes the real part of complex scalar. Then, inspired by the weighted sum mean-square error minimization (WMMSE) algorithm proposed for conventional multiuser MIMO systems \cite{Qingjiang_WMMSE}, we introduce an auxiliary optimization vector variable $\mathbf v=(v_1, v_2,\cdots, v_K)^T$ and establish a more tractable MA-enabled WMMSE problem as follows
\begin{subequations}
\label{p: WMMSE problem}
\begin{align}
\underset{\mathbf W, \mathbf t,\mathbf r,\mathbf u,\mathbf v}{\text{minimize}} \quad& \sum\limits_{k=1}^K \alpha_k (v_k e_k - \log(v_k)) \label{p: WMMSE problem-a}\\
\text { subject to } &\sum\limits_{k=1}^K\|\mathbf w_k\|^2\leq P_\mathrm{max},\label{p: WMMSE problem-b}\\
&\|\mathbf t_i - \mathbf t_j\|_2\geq D, \forall i\neq j,\label{p: WMMSE problem-c}\\
&\mathbf t_i\in\mathcal C^t,\quad\mathbf r_k \in\mathcal C_k^r,\label{p: WMMSE problem-d}\\
&\mathbf v\geq \mathbf 0.\label{p: WMMSE problem-e}
\end{align}
\end{subequations}
\textit{Remark:} The equivalence between \eqref{p: original problem} and \eqref{p: WMMSE problem} can be proved by deriving the optimal $u_k,v_k,\forall k$, in Eq. \eqref{p: WMMSE problem-a} and substituting them back into Eq. \eqref{p: WMMSE problem-a}, which is similar to \cite[Appendix A]{Qingjiang_WMMSE} for the traditional FPA MIMO system.

\section{Block Coordinate Descent Method for MA-Enhanced Multiuser MIMO System}
BCD method decomposes the optimization variables into multiple blocks and optimizes different blocks in each iteration with the other blocks fixed.

\subsection{Transceiver Beamformer Design}
According to the principle of BCD, we first investigate the optimization of beamformer $\mathbf W$ and related variables $\mathbf u, \mathbf v$ for fixed MA position vectors $\mathbf t, \mathbf r$. The beamformer design problem accordingly reduces to
\begin{subequations}
\label{p: beamformer}
\begin{align}
\underset{\mathbf W, \mathbf u,\mathbf v}{\text{minimize}} \quad& \sum\limits_{k=1}^K \alpha_k (v_k e_k - \log(v_k)) \label{p: beamformer-a}\\
\text { subject to } &\sum\limits_{k=1}^K\|\mathbf w_k\|^2\leq P_\mathrm{max},\label{p: beamformer-b}\\
&\mathbf v\geq \mathbf 0.\label{p: beamformer-c}
\end{align}
\end{subequations}
This problem has been studied extensively in the literature, and one of the well-known methods to obtain a stationary solution is the WMMSE algorithm with the following iterative updating rule \cite[Table I]{Qingjiang_WMMSE}:
\begin{subequations}
\begin{align}
u_k &= \left(\sum_{i=1}^K |\mathbf h_k(\mathbf t,\mathbf r_k)^H \mathbf w_i|^2 + \sigma^2\right)^{-1}\mathbf h_k(\mathbf t,\mathbf r_k)^H \mathbf w_k,\label{eq: beamformer-u}\\
v_k &= \left(1 - u_k^*\mathbf h_k(\mathbf t,\mathbf r_k)^H \mathbf w_k\right)^{-1},\label{eq: beamformer-v}\\
\mathbf w_k &= \alpha_k u_k v_k \left(\mu \mathbf I_M + \sum_{i=1}^K \alpha_i|u_i|^2 v_i \mathbf h_i(\mathbf t,\mathbf r_k)\mathbf h_i(\mathbf t,\mathbf r_k)^H\right)^{-1}\notag\\
&\qquad\mathbf h_k(\mathbf t,\mathbf r_k),\label{eq: beamformer-w}
\end{align}
\end{subequations}
where $\mu \geq 0$ is the optimal dual variable for the transmit power constraint in \eqref{p: beamformer-b}. To describe the computation of $\mu$, we let $\mathbf w_k(\mu)$ denote the right-hand side of \eqref{eq: beamformer-w}. When the matrix $\sum_{i=1}^K \alpha_i|u_i|^2 v_i \mathbf h_i(\mathbf t,\mathbf r_k)\mathbf h_i(\mathbf t,\mathbf r_k)^H$ is invertible and $\sum_{k=1}^K \|\mathbf w_k(0)\|^2 \leq P_\mathrm{max}$, then $\mathbf w_k = \mathbf w_k(0)$, otherwise the optimal $\mu$ should satisfy $\sum_{k=1}^K \|\mathbf w_k(\mu)\|^2 = P_\mathrm{max}$. Note that the left-hand side is a decreasing function in $\mu$ for $\mu>0$. Hence, it can be obtained via one dimensional search techniques, e.g., bisection method.

\subsection{MA Position Design at BS}
To reduce the computational complexity, we resort to alternative optimization to design MA positions at the BS. For fixed $\mathbf W$, $\mathbf t_j,\forall j\neq m$, and $\mathbf r$, the problem of designing MA $m$ position at BS accordingly reduces to
\begin{subequations}
\label{p: position}
\begin{align}
\underset{\mathbf t_m\in\mathcal C^t}{\text{minimize}} \,\,\,& \sum_{k=1}^K\left(\mathbf f_k^H(\mathbf t_m) \mathbf A_{k,m} \mathbf f_k(\mathbf t_m) + \mathrm{Re}(\mathbf b_{k,m}^H \mathbf f_k(\mathbf t_m))\right)\\
\text { subject to } &\|\mathbf t_m - \mathbf t_j\|_2\geq D, \forall j\neq m,\label{p: position-c}
\end{align}
\end{subequations}
where
\begin{equation}
\begin{aligned}
\mathbf A_{k,m} &\triangleq \alpha_k v_k |u_k|^2 \|\mathbf w_{m,:}\|^2 \mathbf \Sigma_k\mathbf g_k(\mathbf r_k)\mathbf g_k(\mathbf r_k)^H \mathbf\Sigma_k^H,\\
\mathbf b_{k,m} &\triangleq 2\alpha_k v_k \bigg(|u_k|^2\sum_{j=1}^K\bigg(w_{m,j}^* \sum_{n\neq m}w_{n,j} \\
&\qquad\mathbf g_k(\mathbf r_k)^H \mathbf\Sigma_k^H\mathbf f_k(\mathbf t_n) \bigg) - u_k w_{m,k}^*\bigg)\mathbf\Sigma_k\mathbf g_k(\mathbf r_k).
\end{aligned}
\end{equation}
Next, we propose to leverage the minorization maximization (MM) \cite{MM} technique to update antenna position vector. The key to the success of MM lies in constructing a surrogate function satisfying the upper bound property for both objective function and constraints. Now, we introduce some Lemmas that facilitate the implementation of MM.
\begin{Lemma}
\label{lemma: quadratic}
The quadratic form $\mathbf x^H \mathbf L \mathbf x$, where $\mathbf L$ is a Hermitian matrix, can be upper bounded as \cite[Example 13]{MM}:
\begin{equation}
\begin{aligned}
\mathbf x^H\mathbf L\mathbf x\leq \mathbf x^H\mathbf M\mathbf x+2\mathrm{Re}(\mathbf x^H(\mathbf L-\mathbf M)\mathbf x_0) + \mathbf x_0^H(\mathbf M-\mathbf L)\mathbf x_0,
\end{aligned}
\end{equation}
where $\mathbf M\succeq\mathbf L$. Equality is achieved at $\mathbf x = \mathbf x_0$.
\end{Lemma}
\begin{Lemma}
\label{lemma: linear}
The linear form $z_{k}(\mathbf t_m)=\mathrm{Re}(\mathbf b^H \mathbf f_k(\mathbf t_m))$, where $\mathbf f_k(\mathbf t_m)$ is the field-response vector defined in \eqref{eq: field-response vector}, can be lower bouneded and upper bounded as \cite[Eq. (26) and Appendix A-B]{Wenyan_MIMO}:
\begin{subequations}
\begin{align}
z_{k}(\mathbf t_m) &\geq z_{k}(\mathbf t_{m,0}) + \nabla z_{k}(\mathbf t_{m,0})^T(\mathbf t_m - \mathbf t_{m,0})\notag\\
&-\frac{4\pi^2}{\lambda^2}\|\mathbf b\|_1 (\mathbf t_m - \mathbf t_{m,0})^T (\mathbf t_m - \mathbf t_{m,0}),\label{linear-a}\\
z_{k}(\mathbf t_m) &\leq z_{k}(\mathbf t_{m,0}) + \nabla z_{k}(\mathbf t_{m,0})^T(\mathbf t_m - \mathbf t_{m,0})\notag\\
&+\frac{4\pi^2}{\lambda^2}\|\mathbf b\|_1 (\mathbf t_m - \mathbf t_{m,0})^T (\mathbf t_m - \mathbf t_{m,0})\label{linear-b},
\end{align}
\end{subequations}
where 
\begin{equation}
\begin{aligned}
\nabla z_{k}(\mathbf t_{m,0})&=\mathrm{Re}(\mathbf b^H \nabla \mathbf f_k(\mathbf t_{m,0}))^T,\\
\nabla \mathbf f_k(\mathbf t_{m,0}) &= j\frac{2\pi}{\lambda}(\mathbf n_{1,k},\cdots, \mathbf n_{L_k^t,k})^T \odot (\mathbf f_k(\mathbf t_{m,0}), \mathbf f_k(\mathbf t_{m,0})).
\end{aligned}
\end{equation}
Equality is achieved at $\mathbf t_m = \mathbf t_{m,0}$.
\end{Lemma}
\textit{Remark:} The proof of Eq. \eqref{linear-a} is provided in \cite[Eq. (26) and Appendix A-B]{Wenyan_MIMO}. Thus, we can derive Eq. \eqref{linear-b} directly by replacing $\mathbf b$ with $-\mathbf b$ in \eqref{linear-a}.

From the Cauchy-Schwartz inequality, at iteration $(n)$, a minorizing function for the constraint is constructed at $\mathbf t_m = \mathbf t_m^{(n)}$ as follows:
\begin{equation}
\|\mathbf t_m - \mathbf t_j\|_2 \geq \frac{(\mathbf t_m^{(n)} - \mathbf t_j)^T(\mathbf t_m - \mathbf t_j)}{\|\mathbf t_m^{(n)} - \mathbf t_j\|_2}.
\end{equation}

Based on Lemmas \ref{lemma: quadratic}-\ref{lemma: linear}, at iteration $(n)$, a majorizing function for the objective function is constructed at $\mathbf t_m = \mathbf t_m^{(n)}$ as follows:
\begin{equation}
\label{eq: upper bound of objective}
\begin{aligned}
&\sum_{k=1}^K\left(\mathbf f_k^H(\mathbf t_m) \mathbf A_{k,m} \mathbf f_k(\mathbf t_m) + \mathrm{Re}(\mathbf b_{k,m}^H \mathbf f_k(\mathbf t_m))\right)\\
&\leq \sum_{k=1}^K\mathrm{Re}(\hat{\mathbf b}_{k,m}^{H}\mathbf f_k(\mathbf t_m)) + \mathrm{const.}\\
&\leq \sum_{k=1}^K \frac{4\pi^2}{\lambda^2}\|\hat{\mathbf b}_{k,m}\|_1 \|\mathbf t_m\|^2 \\
&\quad+ (\nabla z_{k}(\mathbf t_{m}^{(n)}) - \frac{8\pi^2}{\lambda^2}\|\hat{\mathbf b}_{k,m}\|_1 \mathbf t_{m}^{(n)})^T\mathbf t_m + \mathrm{const.}
\end{aligned}
\end{equation}
where
\begin{equation}
\begin{aligned}
\hat{\mathbf b}_{k,m} &\triangleq 2(\mathbf A_{k,m} - \alpha_k v_k |u_k|^2 \|\mathbf w_{m,:}\|^2 \\
&\qquad\cdot\|\mathbf \Sigma_k\mathbf g_k(\mathbf r_k)\|^2 \mathbf I)\mathbf f_k(\mathbf t_m^{(n)}) + \mathbf b_{k,m},\\
z_{k}(\mathbf t_m) &\triangleq \mathrm{Re}(\hat{\mathbf b}_{k,m}^{H}\mathbf f_k(\mathbf t_m)).
\end{aligned}
\end{equation}

Therefore, the antenna position vector $\mathbf t_m$ is updated as 
\begin{subequations}
\begin{align}
\mathbf t_m^{(n+1)}&=\underset{\mathbf t_m\in\mathcal C^t}{\text{argmin}}  \sum_{k=1}^K \frac{4\pi^2}{\lambda^2}\|\hat{\mathbf b}_{k,m}\|_1 \|\mathbf t_m\|^2\notag\\
&\qquad\quad+ (\nabla z_{k}(\mathbf t_{m}^{(n)}) - \frac{8\pi^2}{\lambda^2}\|\hat{\mathbf b}_{k,m}\|_1 \mathbf t_{m}^{(n)})^T\mathbf t_m\\
&\text { subject to } \frac{(\mathbf t_m^{(n)} - \mathbf t_j)^T(\mathbf t_m - \mathbf t_j)}{\|\mathbf t_m^{(n)} - \mathbf t_j\|_2}\geq D, \forall j\neq m,
\end{align}
\end{subequations}
which is a typical convex quadratic programming (QP) problem and can be efficiently solved by quadprog or CVX.

\subsection{MA Position Design at Users}
For fixed $\mathbf W$, $\mathbf u$, $\mathbf v$, and $\mathbf t$, each $\mathbf r_k$ can be updated in parallel since they are independent in \eqref{p: WMMSE problem}. Specifically, the problem of designing MA position at user $k$ accordingly reduces to
\begin{equation}
\label{p: position_user}
\begin{aligned}
\underset{\mathbf r_k\in\mathcal C_k^r}{\text{minimize}} \,\,\,& \mathbf g_k^H(\mathbf r_k) \mathbf C_{k} \mathbf g_k(\mathbf r_k) + \mathrm{Re}(\mathbf d_{k}^H \mathbf g_k(\mathbf r_k))
\end{aligned}
\end{equation}
where
\begin{equation}
\begin{aligned}
\mathbf C_{k} &\triangleq \sum_{j=1}^K |u_k|^2 \mathbf\Sigma_k^H \mathbf F_k(\mathbf t) \mathbf w_j \mathbf w_j^H \mathbf F_k(\mathbf t)^H  \mathbf \Sigma_k,\\
\mathbf d_{k} &\triangleq -2u_k^* \mathbf\Sigma_k^H \mathbf F_k(\mathbf t) \mathbf w_k.
\end{aligned}
\end{equation}

Similar to Lemmas \ref{lemma: quadratic}-\ref{lemma: linear}, at iteration $(n)$, a majorizing function for the objective function is constructed at $\mathbf r_k = \mathbf r_k^{(n)}$ as follows:
\begin{equation}
\begin{aligned}
&\mathbf g_k^H(\mathbf r_k) \mathbf C_{k} \mathbf g_k(\mathbf r_k) + \mathrm{Re}(\mathbf d_{k}^H \mathbf g_k(\mathbf r_k))\\
&\leq \frac{4\pi^2}{\lambda^2}\|\hat{\mathbf d}_{k}\|_1 \|\mathbf r_k\|^2 \\
&\quad+ (\nabla z_{k}(\mathbf r_{k}^{(n)}) - \frac{8\pi^2}{\lambda^2}\|\hat{\mathbf d}_{k}\|_1 \mathbf r_{k}^{(n)})^T\mathbf r_k + \mathrm{const.}
\end{aligned}
\end{equation}
where
\begin{equation}
\begin{aligned}
\hat{\mathbf d}_{k} &\triangleq 2(\mathbf C_{k} - \sum_{j=1}^K |u_k|^2 \|\mathbf\Sigma_k^H \mathbf F_k(\mathbf t) \mathbf w_j\|^2 \mathbf I)\mathbf g_k(\mathbf r_k^{(n)}) + \mathbf d_{k}\\
z_{k}(\mathbf r_k) &\triangleq \mathrm{Re}(\hat{\mathbf d}_{k}^{H}\mathbf g_k(\mathbf r_k)).
\end{aligned}
\end{equation}

Therefore, the MA position at user $k$ is updated as 
\begin{equation}
\begin{aligned}
&\mathbf r_k^{(n+1)}=\underset{\mathbf r_k\in\mathcal C_k^r}{\text{argmin}}  \frac{4\pi^2}{\lambda^2}\|\hat{\mathbf d}_{k}\|_1 \|\mathbf r_k\|^2\\
&\qquad\qquad\qquad+ (\nabla z_{k}(\mathbf r_{k}^{(n)}) - \frac{8\pi^2}{\lambda^2}\|\hat{\mathbf d}_{k}\|_1 \mathbf r_{k}^{(n)})^T\mathbf r_k\\
&=\Pi_{\mathcal C_k^r}\left(\mathbf r_{k}^{(n)} - \frac{\sum_{k=1}^K \nabla z_{k}(\mathbf r_{k}^{(n)})}{\sum_{k=1}^K \frac{8\pi^2}{\lambda^2}\|\hat{\mathbf d}_{k}\|_1}\right).
\end{aligned}
\end{equation}

\section{Low-Complexity Design}
To reduce the computational complexity, we propose the planar movement mode at BS, where each MA is only allowed to move in given planar area and the minimum distance between any two areas is set as $D$ to avoid coupling effect. To distinguish, the movement mode described in Section III is referred to as the general movement mode.

\subsection{Planar Movement Mode at BS}
The position optimization problem \eqref{p: position} can be recast as
\begin{equation}
\begin{aligned}
\underset{\mathbf t_m\in\mathcal C_m^t}{\text{minimize}} \,\,\,& \sum_{k=1}^K\left(\mathbf f_k^H(\mathbf t_m) \mathbf A_{k,m} \mathbf f_k(\mathbf t_m) + \mathrm{Re}(\mathbf b_{k,m}^H \mathbf f_k(\mathbf t_m))\right),
\end{aligned}
\end{equation}
where $\mathcal C_m^t$ is the movement area of MA $m$. Therefore, through constructing the tight upper bound of the objective function as Eq. \eqref{eq: upper bound of objective}, the antenna position vector $\mathbf t_m$ is updated as 
\begin{equation}
\begin{aligned}
&\mathbf t_m^{(n+1)}=\underset{\mathbf t_m\in\mathcal C_m^t}{\text{argmin}}  \sum_{k=1}^K \frac{4\pi^2}{\lambda^2}\|\hat{\mathbf b}_{k,m}\|_1 \|\mathbf t_m\|^2\\
&\qquad\qquad\qquad+ (\nabla z_{k}(\mathbf t_{m}^{(n)}) - \frac{8\pi^2}{\lambda^2}\|\hat{\mathbf b}_{k,m}\|_1 \mathbf t_{m}^{(n)})^T\mathbf t_m\\
&=\Pi_{\mathcal C_m^t}\left(\mathbf t_{m}^{(n)} - \frac{\sum_{k=1}^K \nabla z_{k}(\mathbf t_{m}^{(n)})}{\sum_{k=1}^K \frac{8\pi^2}{\lambda^2}\|\hat{\mathbf b}_{k,m}\|_1}\right).
\end{aligned}
\end{equation}

\subsection{Convergence and Complexity Analysis}
\emph{Convergence analysis:} Based on the WMMSE algorithm for transmit beamformer design and the MM algorithm for MA position design, a stationary point is guaranteed for each subproblem \cite{Qingjiang_WMMSE,MM}. Therefore, the initial equivalent function is guaranteed to increase monotonically, and our proposed algorithm is convergent.

\emph{Complexity analysis:} The complexity of updating $\mathbf w$ at each iteration of the subproblem is $\mathcal O(K^2 M^2)$ \cite[\S IV]{Qingjiang_WMMSE}. The complexity of updating $\mathbf t$ mainly depends on computing coefficients $\hat{\mathbf b}_{k,m}$ and $\nabla z_{k}(\mathbf t_{m}^{(n)})$ and solving the QP problem. Therefore, the complexity of the general movement mode is $\mathcal O(K M^2 + KML_tL_r + KML_t^2 + M^{2.5})$, while for the planar movement mode, it is $\mathcal O(K M^2 + KML_tL_r + KML_t^2 + M)$ at each iteration of subproblem. Similarly, the complexity of updating $\mathbf r$ is $\mathcal O(K^2 M + KML_tL_r + KML_r^2 + K)$.

\section{Numerical Results}
In this section, we numerically evaluate the performance of the proposed algorithm. The number of MAs at BS is $M=16$. Each element of the path-response matrix is assumed to be independently and identically distributed (i.i.d.) Gaussian. The noise power is set as $\sigma^2 = 15$ dBm \cite{Yuqi_FAS_Statistical_CSI}. The normalized wavelength is set as $\lambda = 1$ m. The minimum distance between adjacent antennas is set as $D=\lambda/2$. The size of the MA movement areas at BS is set as $5\lambda\times5\lambda$.

We compare the proposed design, named TMA-RMA, with the following baselines: i) Traditional antennas: All antenna positions are fixed; ii) TFPA-RMA: The antenna positions at the BS are fixed, but the antenna positions at users can be adjustable in the receive region; iii) TMA-RFPA: The antenna positions at the users are fixed, but the antenna positions at the BS are adjustable in the transmit region.

\begin{figure*}
\begin{minipage}{0.32\textwidth}
\centering
\includegraphics[width=0.9\linewidth]{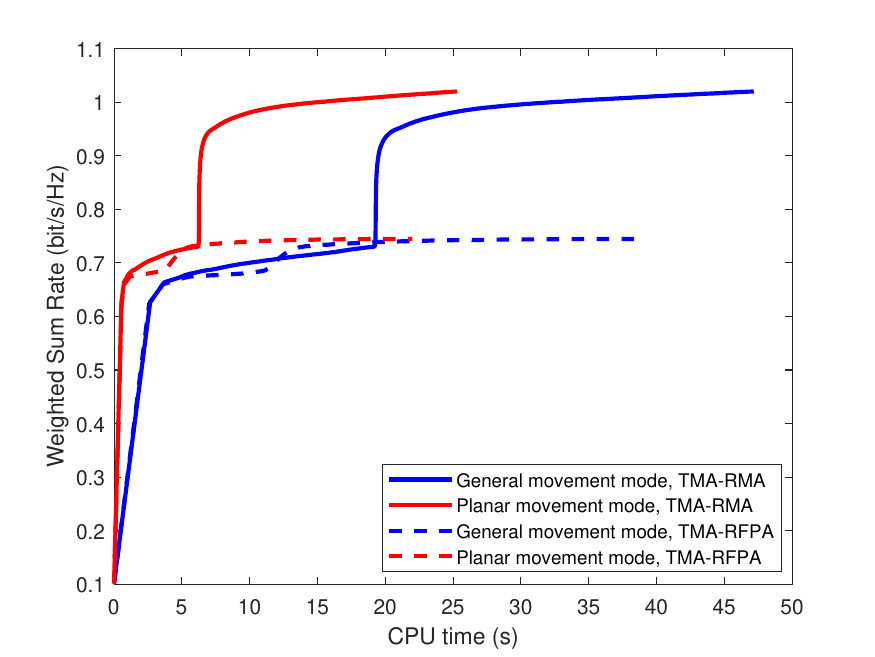}
\caption{Convergence of the proposed algorithms.}
\label{CPUtime}
\end{minipage}
\hfill
\begin{minipage}{0.32\textwidth}
\centering
\includegraphics[width=0.9\linewidth]{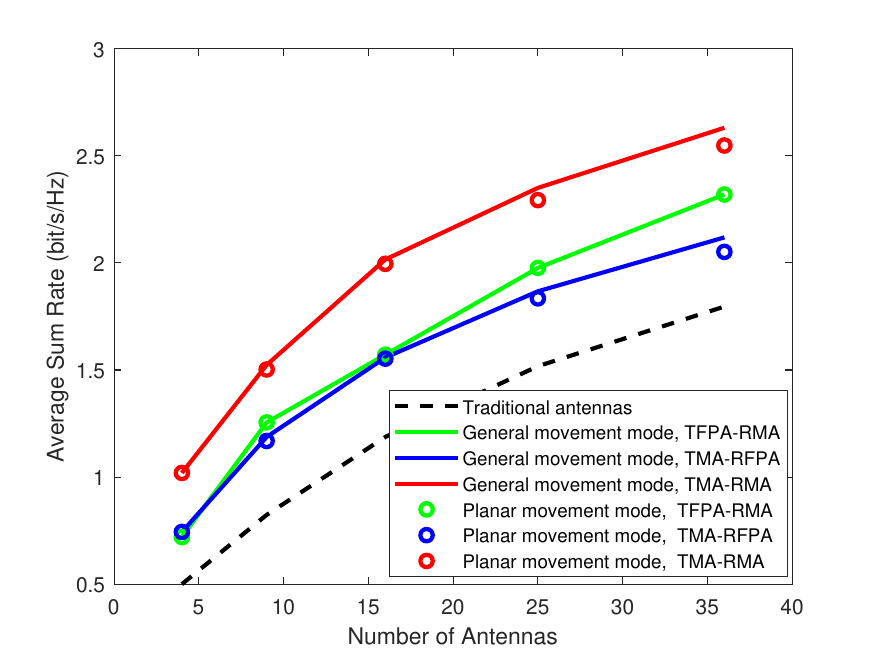}
\caption{Evaluation of weighted sum rate under different number of antennas $M$.}
\label{Power}
\end{minipage}
\hfill
\begin{minipage}{0.32\textwidth}
\centering
\includegraphics[width=0.9\linewidth]{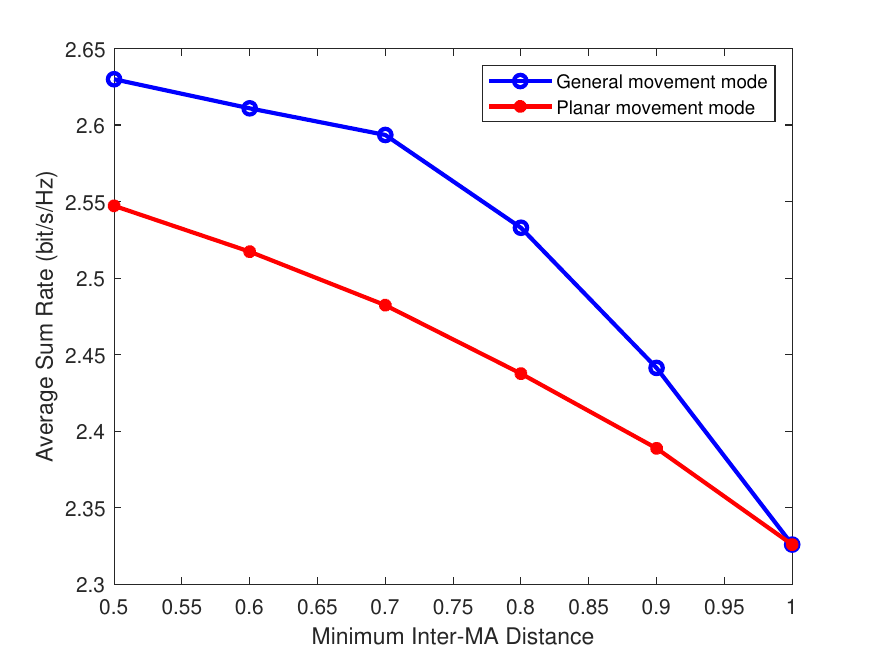}
\caption{Evaluation of weighted sum rate under different minimum inter-MA distance $D$.}
\label{Distance}
\end{minipage}
\end{figure*}

Fig.~\ref{CPUtime} demonstrates the convergence performance of our proposed algorithms. Firstly, it can be observed that the differences in the achieved average sum rate between the two movement modes are relatively small. Secondly, the low-complexity planar movement mode clearly exhibits a faster convergence speed, which aligns with our analysis.

Fig.~\ref{Power} illustrates the average sum rate for various antenna configurations at both BS and users. Firstly, for different numbers of MAs at BS, the proposed design consistently outperforms three baselines in terms of the average sum rate. The performance of the system improves as the number of antennas increases. Secondly, as the DoF of antenna positions increases, so does the system gain. It indicates that deploying MAs holds great promise for enhancing wireless communication systems, especially when compared to the traditional antennas. Finally, the planar movement mode achieves close performance to the general movement mode, which motivates us for the further design of low-complexity systems.

Fig.~\ref{Distance} depicts the average sum rate for different minimum inter-MA distances at BS. As the minimum distance increases, both the DoF and system gain decrease accordingly.

\section{Conclusion}
In this letter, we employed the MAs at both BS and users to improve the weighted sum rate of wireless networks. We adopted BCD algorithm to optimize all variables, and then confined each antenna to a designated area, i.e., planar movement mode, to reduce the computational complexity. Simulation results confirmed the huge potential of MAs in improving the performance of future communications systems.

\end{document}